\documentstyle[12pt,psfig]{article}

\def\bra#1{{\langle#1|}}
\def\ket#1{{|#1\rangle}}

\def\expect#1{{\langle#1\rangle}}
\def\e{{\rm e}}

\def\tr{{\rm Tr}}
\def\H{{\hat H}}

\def\L{{\hat L}}
\def\Ldag{{\hat L}^\dagger}
\def\a{{\hat a}}
\def\adag{{\hat a}^\dagger}
\def\b{{\hat b}}
\def\bdag{{\hat b}^\dagger}
\def\U{{\hat U}}
\def\Udag{{\hat U}^\dagger}

\def\id{{\hat I}}
\def\Z{{\hat Z}}
\def\p{{\hat p}}
\def\Sx{{\hat S}_x}
\def\Sy{{\hat S}_y}
\def\Sz{{\hat S}_z}
\def\Sl{{\hat S}_L}

\begin{document}

\title{Realistic simulations of \\
single-spin nondemolition measurement \\
by magnetic resonance force microscopy}

\author{Todd A. Brun \\
Institute for Advanced Study, Einstein Drive, \\
Princeton, NJ  08540 USA \\ \\
Hsi-Sheng Goan \\
Center for Quantum Computer Technology, \\
University of New South Wales, \\
Sydney, NSW 2052 Australia}

\maketitle

\begin{abstract}
A requirement for many quantum computation schemes is the ability
to measure single spins.  This paper examines one proposed scheme:
magnetic resonance force microscopy, including the effects of thermal
noise and back-action from monitoring.  We derive a simplified equation
using the adiabatic approximation, and produce a stochastic pure state
unraveling which is useful for numerical simulations.
\end{abstract}

\section{Introduction}

Single-spin measurement is an extremely important challenge, and
necessary for the future successful development of
several recent spin-based proposals for quantum
information processing. \cite{Loss98,Kane98,Yablonovitch99,Berman00a,Twamley02}
There are both direct and indirect single-spin measurement proposals. 
The idea behind some indirect proposals is to transform the
problem of detecting a single spin into the task of
measuring charge transport \cite{Kane98,Loss02},
since the ability to detect a single charge is now available.  
For direct single-spin detection, magnetic resonance force microscopy
(MRFM) has been suggested \cite{Sidles91,Sidles92,Berman01b}
as one of the most promising techniques.  To date, the MRFM technique
has been demonstrated with sensitivity to a few hundred spins
\cite{Rugar98,Rugar01}.

In this paper we discuss how to
read out the quantum state of a single spin using the MRFM technique
based on cyclic adiabatic inversion (CAI). \cite{Rugar94,Rugar98,Berman01b}
In this CAI MRFM technique, the frequency of the spin inversion in the
rotating frame is in resonance with the mechanical vibration of an
ultra thin cantilever, allowing it to amplify the otherwise
extremely weak force due to the spin.  These amplified vibrations
can then be detected by, e.g., optical methods.

Previous studies \cite{Sidles92,Berman01b} of the dynamics of
single-spin measurement by MRFM considered only the unitary
evolution of the spin and the cantilever system, without
including any effects of external environments or measurement devices.  
Only recently, the effect of thermal noise
environment on the dynamics of the spin-cantilever system in the MRFM was
studied  \cite{Berman02} 
by using the Caldeira-Leggett master equation \cite{Leggett83}
in the high-temperature limit.   

There is, however, a macroscopic device in the MRFM setup which measures
the cantilever motion and hence provides information about the spin
state.  To our knowledge, 
the back-action of the measurement device and the effect of the thermal noise
on the dynamics of the cantilever-spin system 
for the single-spin detection problem by MRFM
have not yet been investigated systematically.   
In this paper, we include, in our analysis,
a measurement device (a fiber-optic interferometer) to monitor the
position of the cantilever. 
We consider various relevant sources of noise
and calculate the signal-to-noise ratio of the output photocurrent of
the measurement device. We also develop a realistic continuous
measurement model, and discuss the approximations and conditions to
achieve a quantum non-demolition measurement of a single spin by MRFM.
Finally, we present some simulation results of the dynamics of
the single-spin measurement process.

\section{The measurement scheme}

\begin{figure}[htbp]
\centerline{\psfig{file=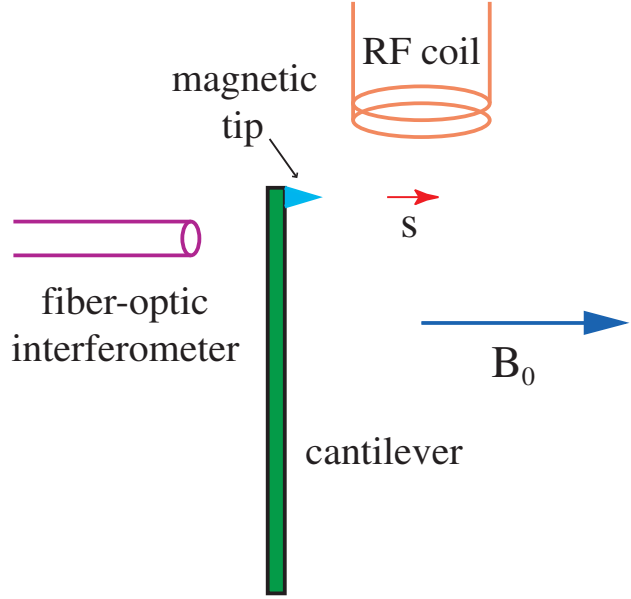,width=0.6\linewidth,angle=0}}
\caption{Schematic diagram of the MRFM setup.}
\label{fig:mrfm_setup}
\end{figure}

A schematic illustration of the MRFM setup is shown in 
Fig \ref{fig:mrfm_setup}.
A uniform magnetic field, $B_0$, points in the positive
$z$-direction.
A single spin is placed in front of the cantilever tip
which can oscillate only in the $z$-direction. 
A ferromagnetic particle (or small magnetic
material) mounted on the cantilever tip produces a non-uniform
magnetic field or magnetic field gradient of 
$({\partial B_z}/{\partial Z})_0$ on the single spin.
As a result, a reactive force (or interaction) acts back on the
magnetic cantilever tip in the $z$-direction from the single spin.
The origin is chosen to be the equilibrium position of the cantilever
tip without the presence of the spin.

In CAI, the cantilever is driven at its resonance frequency 
to amplify the otherwise very small vibrational amplitude.
This is achieved by a modulation scheme using the frequency modulation
of a rotating radio-frequency (RF) magnetic field in the $x$-$y$
plane. In this case, the rotating RF field can be represented as 
$B_{1x}=B_1\cos\{[\omega +\Delta\omega(t)]t\}$,   
$B_{1y}=-B_1\sin\{[\omega +\Delta\omega(t)] t\}$, where
the frequency modulation $\Delta\omega(t)$ is a
periodic function in time with the resonant frequency $\omega_m$ 
of the cantilever.  
In the reference frame rotating with the ${\mathbf B}_1$, the
spin-cantilever Hamiltonian can be written as 
\begin{equation}
\H_{SZ}(t) = \H_Z 
-\hbar\left[\omega_L-\omega-\Delta\omega(t)\right]\Sz
-\hbar\omega_1\Sx-g\mu\left({{\partial B_z}\over{\partial
      Z}}\right)_0 \Z\Sz\;,
\label{Hsc}
\end{equation}
where $\omega_L=g\mu B_z/\hbar$ and $\omega_1=g\mu B_1/\hbar$ are
respectively the Larmor and Rabi frequencies, $B_z$ includes the uniform
magnetic field $B_0$ and the magnetic field produced by the
ferromagnetic particle, $g$ and $\mu$ are the $g$-factor and the
electron or nuclear magneton, and 
\begin{equation}
  \label{eq:HZ}
  \H_Z = \frac{1}{2m}\p^2 + \frac{m\omega_m^2}{2} \Z^2
\end{equation}
is the Hamiltonian of the cantilever in isolation (i.e.,
with no external magnetic field coupling it to the spin).
For $\omega=\omega_L$, we arrive at 
an effective cantilever-spin Hamiltonian of the form
\begin{equation}
\H_{SZ}(t) = \H_Z - 2 \eta \Z \Sz + f(t) \Sz - \varepsilon \Sx \;,
\label{eq:HSZ}
\end{equation}
where $f(t)=\Delta\omega(t)$,
$\eta=(g\mu/2)({\partial B_z}/{\partial Z})_0$
and $\varepsilon=\hbar\omega_1$.
We will discuss in details the rotating picture and adiabatic
approximation for the spin-cantilever system in the next section.

In the following, we briefly describe the basic principle of the 
single-spin measurement by CAI MRFM.   
In the case when the adiabatic approximation is exact, 
the instantaneous eigenstates of the spin
Hamiltonian in the rotating frame of the ${\mathbf B}_1$ field 
are the spin states parallel or antiparallel
to the direction of the effective magnetic field
${\mathbf B}^{\rm eff}(t)=\left(\varepsilon, 0,-f(t) \right)$,
denoted as $|v_\pm(t)\rangle$, respectively.  We define an
operator $\Sz'$ for the component of spin along this axis.
Note that the initial spin state in the laboratory frame has 
the same expression as the initial state in the rotating frame. 
Starting at a general initial spin state (in the laboratory or
rotating frame) of
\begin{equation}
\chi(0)=a|\uparrow\rangle +b|\downarrow\rangle
\end{equation} 
in the $\Sz$ representation,
we can rewrite this initial state in the basis of the instantaneous 
eigenstates of $\Sz'$ as 
\begin{equation}
\chi(0)=a_{\rm eff}|v_+(0)\rangle +
b_{\rm eff}|v_-(0)\rangle,
\end{equation} 
where 
\begin{eqnarray}
a_{\rm eff}&=&a\cos(\Theta_0/2)+b\sin(\Theta_0/2),
\\
b_{\rm eff}&=&-a\sin(\Theta_0/2)+b\cos(\Theta_0/2), 
\end{eqnarray}
and $\Theta_0\equiv \Theta(0)$ 
is the initial angle between the effective magnetic 
magnetic field and the $z$-axis direction. This implies
$\tan[\Theta(t)] = B_x^{\rm eff}(t)/B_z^{\rm eff}(t)=-\varepsilon/f(t)$.
It then follows from the adiabatic theorem that the spin state at time
$t$ can be written as:
\begin{eqnarray}
\chi(t)&=&a_{\rm eff}|v_+(t)\rangle\,\exp({-\frac{i}{\hbar}\int_0^t
\lambda_{+}(t') dt'})
\nonumber \\
&&+
b_{\rm eff}|v_-(t)\rangle\, 
\exp({-\frac{i}{\hbar} \int_0^t \lambda_{-}(t') dt'}),
\end{eqnarray} 
where $\lambda_{\pm}(t)$ are instantaneous eigenvalues.
So the probabilities of finding the spin to be
in the instantaneous eigenstates $|v_\pm(t)\rangle$
are respectively $|a_{\rm eff}|^2$ and $|b_{\rm eff}|^2$.
Since the coefficients $a_{\rm eff}$ and $b_{\rm eff}$ 
are time independent, the probabilities 
$|a_{\rm eff}|^2$ and $|b_{\rm eff}|^2$
remain the same at all times.  This provides us with an opportunity
to measure the initial spin state probabilities at later times.

How do we measure these spin state probabilities?
The idea is to transfer the information of the spin
state to the state of the driven cantilever. 
In the interaction picture in which the state is
rotating with the instantaneous eigenstates of the
spin Hamiltonian, the spin-cantilever
interaction can be written as $2\eta \Z \Sz' \cos[\Theta(t)]$.
As a result, the phase of the driven cantilever vibrations
depends on the orientation of the spin states.
Suppose that the initial state is a product state of the
cantilever and spin parts. 
At a later time, due to the interaction between them,
the total state becomes entangled.
Monitoring the phase of the cantilever vibrations
will give us the information about the spin.  
Numerical simulations (see Fig.\ \ref{fig:z_vs_t_phase} 
with reasonable parameters for the
CAI approximations) indicate that as the amplitude of the
cantilever vibrations increases with time, the phase 
difference in the oscillations for the two different initial
spin eigenstates of $\Sz'$ approaches $\pi$.    
In other words, the measurement of the single-spin states can be achieved  
by monitoring the phases of the cantilever vibrations
at some later time $t$. Phase-sensitive, optical homodyne measurements of
the cantilever vibrations can be performed
using a fiber-optic interferometer.
The main purpose of this paper is to present  
a realistic and detailed analysis of the single-spin measurement
scheme, including the effects of the measurement device and 
other relevant sources of noise.

\section{The rotating picture and adiabatic approximation}

We assume an effective cantilever-spin Hamiltonian of the form (\ref{eq:HSZ})
where for the moment we let $f(t)$ and $\varepsilon$ be arbitrary,
and $\H_Z$ is the Hamiltonian given by (\ref{eq:HZ}).  It is
useful to group this into three terms
\begin{equation}
\H_{SZ}(t) = \H_Z + \H_I + \H_S(t) \;,
\end{equation}
where
\begin{eqnarray}
\H_I &\equiv& - 2 \eta \Z \Sz \;, \nonumber\\
\H_S(t) &\equiv& f(t) \Sz - \varepsilon \Sx \;.
\end{eqnarray}
The state of the cantilever-spin system evolves according to
the Schr\"odinger equation
\begin{equation}
\frac{d\ket{\psi(t)}}{dt} = - \frac{i}{\hbar} \H_{SZ}(t) \ket{\psi(t)} \;.
\label{schrodinger}
\end{equation}

In realistic cases, the spin part of the Hamiltonian (representing
precession under the magnetic field) gives an evolution which is very
rapid compared to the reaction time of the cantilever.  It therefore
makes sense to switch to an interaction picture in which the state
is rotating along with this precession.  We do this by introducing
a (partial) time translation operator
\begin{equation}
\U_S(t) \equiv
  : \exp - \frac{i}{\hbar}\left[ \int_0^t \H_S(t') dt' \right] : \;,
\end{equation}
where $::$ indicates that the integral is to be taken in a time-ordered
sense; this unitary operator obeys the differential equation
\begin{equation}
\frac{d\U_S(t)}{dt} = - \frac{i}{\hbar} \H_S(t) \U_S(t) \;.
\label{unitary_evolution}
\end{equation}

We then introduce the state $\ket{\tilde\psi(t)}$ in the {\it rotating}
picture:
\begin{equation}
\ket{\tilde\psi(t)} \equiv \Udag_S(t) \ket{\psi(t)} \;,
\end{equation}
with $\ket{\psi(t)}$ the solution of the original Schr\"odinger equation
(\ref{schrodinger}) at time $t$.  The evolution equation for
$\ket{\tilde\psi(t)}$ is
\begin{eqnarray}
\frac{d\ket{\tilde\psi(t)}}{dt} &=&
  \frac{d\Udag_S(t)}{dt} \ket{\psi(t)} +
  \Udag_S(t) \frac{d\ket{\psi(t)}}{dt} \nonumber\\
&=& \frac{i}{\hbar} \Udag_S(t) \H_S(t) \ket{\psi(t)} -
  \frac{i}{\hbar} \Udag_S(t) \H_{SZ}(t) \ket{\psi(t)} \nonumber\\
&=& - \frac{i}{\hbar} \H_Z \ket{\tilde\psi(t)}
  - \frac{i}{\hbar}
  \left[ \Udag_S(t) \H_I \U_S(t) \right] \ket{\tilde\psi(t)} \nonumber\\
&=& - \frac{i}{\hbar} \H_Z \ket{\tilde\psi(t)}
  + \frac{2 i \eta}{\hbar}
  \Z \left[ \Udag_S(t) \Sz \U_S(t) \right] \ket{\tilde\psi(t)} \;.
\label{rotating}
\end{eqnarray}
We can define a {\it locked spin} operator $\Sl(t)$
\begin{equation}
\Sl(t) \equiv \left[ \Udag_S(t) \Sz \U_S(t) \right] \;;
\end{equation}
in terms of this, the equation of motion for $\ket{\tilde\psi}$ becomes
\begin{equation}
\frac{d\ket{\tilde\psi(t)}}{dt} =
  - \frac{i}{\hbar} \H_Z \ket{\tilde\psi(t)}
  + \frac{2 i \eta}{\hbar} \Z \Sl(t) \ket{\tilde\psi(t)} \;.
\end{equation}

Unfortunately, it is difficult to get an exact solution for
$\U_S(t)$ for a general function $f(t)$.  This means that it is also
difficult to derive an exact expression for $\Sl(t)$, and the
rotating picture (\ref{rotating}), while formally correct, is not
very helpful.

However, while we cannot easily find an exact expression for
$\U_S(t)$ for general $f(t)$, we can easily find an {\it approximate}
solution for a large class of functions.  Suppose that
$\varepsilon$ is large and $f(t)$ is slowly varying,
so that $|f(t)|,\varepsilon \gg |f'(t)/f(t)|$ for typical values
of $f(t)$ and $f'(t)$.  Then $\H_S(t)$ is also slowly varying, and if
a spin begins in an {\it instantaneous eigenstate} of $\H_S(t)$, it will
remain close to an instantaneous eigenstate of $\H_S(t)$ for all
times by the adiabatic theorem.

The instantaneous eigenstates of $\H_S(t)$ are
\begin{equation}
\H_S(t) \ket{v_\pm(t)} = \lambda_\pm(t) \ket{v_\pm(t)} \equiv
  \pm \lambda(t) \ket{v_\pm(t)} \;,
\end{equation}
where
\begin{eqnarray}
\lambda(t) &=& \sqrt{f^2(t) + \varepsilon^2} \;, \nonumber\\
\ket{v_\pm(t)} &=&
  \frac{\varepsilon}{\sqrt{(f(t) \mp \lambda(t))^2 + \varepsilon^2}}
  \ket\downarrow
- \frac{f(t)\mp\lambda(t)}{\sqrt{(f(t) \mp \lambda(t))^2 + \varepsilon^2}}
  \ket\uparrow \;.
\label{eigensystem}
\end{eqnarray}
We use these instantaneous eigenvectors and eigenvalues to define
an approximation to the unitary operator $\U_S(t)$:
\begin{equation}
\U'_S(t) = \id \otimes \ket{v_+(t)}\bra{v_+(0)} \e^{-i\Phi(t)}
 + \id \otimes \ket{v_-(t)}\bra{v_-(0)} \e^{i\Phi(t)} \;,
\label{approx_unitary}
\end{equation}
with the accumulated phase
\begin{equation}
\Phi(t) \equiv \frac{1}{\hbar} \int_0^t \lambda(t') dt' \;.
\end{equation}

Note that $\Phi(t)$ obeys $d\Phi(t)/dt = \lambda(t)$.  This implies that
\begin{eqnarray}
\frac{d\U'_S(t)}{dt} &=&
 - \frac{i}{\hbar}\lambda(t)
  \id \otimes \ket{v_+(t)}\bra{v_+(0)} \e^{-i\Phi(t)}
 + \frac{i}{\hbar}\lambda(t)
  \id \otimes \ket{v_-(t)}\bra{v_-(0)} \e^{i\Phi(t)} \nonumber\\
&& + \id \otimes \frac{d\ket{v_+(t)}}{dt}\bra{v_+(0)} \e^{-i\Phi(t)}
  + \id \otimes \frac{d\ket{v_-(t)}}{dt}\bra{v_-(0)} \e^{i\Phi(t)} \nonumber\\
&=& - \frac{i}{\hbar} \H_S(t) \U'_S(t)
  + \id \otimes \frac{d\ket{v_+(t)}}{dt}\bra{v_+(0)} \e^{-i\Phi(t)} \nonumber\\
&& + \id \otimes \frac{d\ket{v_-(t)}}{dt}\bra{v_-(0)} \e^{i\Phi(t)} \;,
\label{unitary_evolution2}
\end{eqnarray}
which has the form of (\ref{unitary_evolution}) plus some additional
terms.  From the definition (\ref{eigensystem}) of $\ket{v_\pm(t)}$, we see
\begin{equation}
\frac{d\ket{v_\pm(t)}}{dt} = \pm \frac{1}{2}
  \frac{\varepsilon}{\lambda^2(t)} \frac{df(t)}{dt} \ket{v_\mp(t)} \;.
\end{equation}
Provided that $f(t)$ is slowly varying, the additional terms in
(\ref{unitary_evolution2}) will be small.

Just as before, we can define a rotating picture, now using the
unitary transformation $\U'_S(t)$,
\begin{equation}
\ket{\breve\psi(t)} \equiv \left(\U'_S(t)\right)^\dagger \ket{\psi(t)} \;.
\end{equation}
This gives us a new evolution equation for $\ket{\breve\psi}$:
\begin{eqnarray}
\frac{d\ket{\breve\psi(t)}}{dt} &=&
  \frac{d(\U'_S(t))^\dagger}{dt} \ket{\psi(t)} +
  (\U'_S(t))^\dagger \frac{d\ket{\psi(t)}}{dt} \nonumber\\
&=& - \frac{i}{\hbar} \H_Z \ket{\breve\psi(t)}
  + \frac{2 i \eta}{\hbar} \Z \left[ (\U'_S(t))^\dagger \Sz \U'_S(t) \right]
  \ket{\breve\psi(t)} \nonumber\\
&& + \id \otimes \Biggl( \ket{v_+(0)}\frac{d\bra{v_+(t)}}{dt} \e^{-i\Phi(t)}
  \nonumber\\
&& + \ket{v_-(0)}\frac{d\bra{v_-(t)}}{dt} \e^{i\Phi(t)} \Biggr)
  \U'_S(t) \ket{\breve\psi(t)} \;.
\label{rotating2}
\end{eqnarray}

At this point, it is helpful to introduce a new set of spin operators
\begin{eqnarray}
\Sx' &=& \frac{\hbar}{2} \id \otimes \left( \ket{v_+(0)}\bra{v_-(0)}
  + \ket{v_-(0)}\bra{v_+(0)} \right) \;, \nonumber\\
\Sy' &=& \frac{i\hbar}{2} \id \otimes \left( \ket{v_-(0)}\bra{v_+(0)}
  - \ket{v_+(0)}\bra{v_-(0)} \right) \;, \nonumber\\
\Sz' &=& \frac{\hbar}{2} \id \otimes \left( \ket{v_+(0)}\bra{v_+(0)}
  - \ket{v_-(0)}\bra{v_-(0)} \right) \;.
\label{spin_prime}
\end{eqnarray}
Using the definition (\ref{approx_unitary}) for $\U'_S(t)$, we can
solve for the various terms in (\ref{rotating2}):
\begin{equation}
(\U'_S(t))^\dagger \Sz \U'_S(t) =
  - \frac{f(t)}{\lambda(t)} \Sz'
  - \frac{\varepsilon}{\lambda(t)}
  \left( \Sx' \cos(2\Phi(t))
  - \Sy' \sin(2\Phi(t)) \right) \;.
\label{locked_spin2}
\end{equation}
\begin{eqnarray}
\id \otimes \left( \ket{v_+(0)}\frac{d\bra{v_+(t)}}{dt} \e^{-i\Phi(t)}
  + \ket{v_-(0)}\frac{d\bra{v_-(t)}}{dt} \e^{i\Phi(t)} \right) \U'_S(t)
  = \ \ \  \nonumber\\
  \frac{i\varepsilon}{\hbar\lambda^2(t)} \frac{df(t)}{dt}
  \left( \Sx' \sin(2\Phi(t))
  + \Sy' \cos(2\Phi(t)) \right) \;.
\label{osc_terms}
\end{eqnarray}
Substituting (\ref{spin_prime}--\ref{osc_terms}) into (\ref{rotating2}),
we get
\begin{eqnarray}
\frac{d\ket{\breve\psi(t)}}{dt} &=&
  - \frac{i}{\hbar} \H_Z \ket{\breve\psi(t)}
  + \frac{2 i \eta}{\hbar}
  \Z \frac{f(t)}{\lambda(t)} \Sz' \ket{\breve\psi(t)} \nonumber\\
&& + \frac{2 i \eta}{\hbar} \Z \frac{\varepsilon}{\lambda(t)}
  \left( \Sx' \cos(2\Phi(t)) - \Sy' \sin(2\Phi(t)) \right)
  \ket{\breve\psi(t)} \nonumber\\
&& +  i \frac{\varepsilon}{\hbar\lambda^2(t)} \frac{df(t)}{dt}
  \left( \Sx' \sin(2\Phi(t)) + \Sy' \cos(2\Phi(t)) \right)
  \ket{\breve\psi(t)} \;.
\label{rotating3}
\end{eqnarray}

Note that this equation is still exact---it is equivalent to the original
Schr\"odinger equation (\ref{schrodinger}).  However, we can see that
if $|f(t)|,\varepsilon$ are large, then $\Phi(t)$ will be a rapidly growing
function, and the last two terms of equation (\ref{rotating3}) will
oscillate very rapidly compared to the first two terms.  Over a short
period relative to the response time of the cantilever they will
essentially average away to nothing.  In this limit, therefore, we
can reasonably make a rotating-wave approximation, to get the approximate
evolution equation
\begin{equation}
\frac{d\ket{\breve\psi(t)}}{dt} \approx
  - \frac{i}{\hbar} \left( \H_Z - 2 \eta (f(t)/\lambda(t)) \Z \Sz' \right)
  \ket{\breve\psi(t)} \;.
\label{rotating_approx}
\end{equation}
This is equivalent to making an exact adiabatic approximation, as
described in section 2.  We can see how this approximation compares
to the complete Hamiltonian for a reasonable set of parameter values
in figures 2 and 3.  This set of parameters was chosen to match those
of Berman et al. \cite{Berman01b}, as was the set of times plotted
in figure 3.  Comparison shows that our results match their unitary
simulations to good precision.  For the rest of this paper we will be using
the rotating wave approximation, and representing states in the
rotating frame.  For simplicity, we henceforth omit the accent
from the state $\ket{\breve\psi}$.

\begin{figure}[htbp]
\centerline{\psfig{file=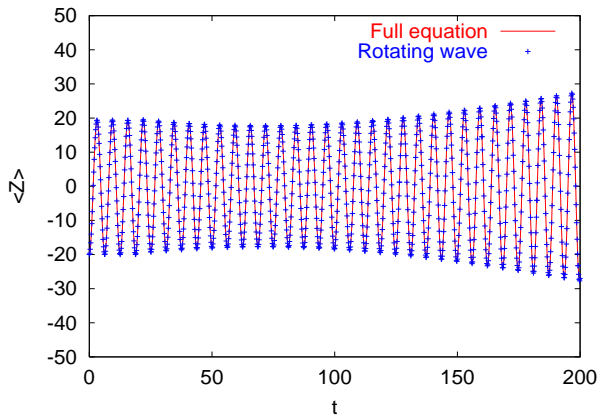,width=0.6\linewidth,angle=0}}
\caption{Mean cantilever position $\expect{\Z}$ vs. $t$ for the complete
and rotating wave Hamiltonians.}
\label{fig:z_vs_t_approx}
\end{figure}

\begin{figure}[htbp]
\centerline{\psfig{file=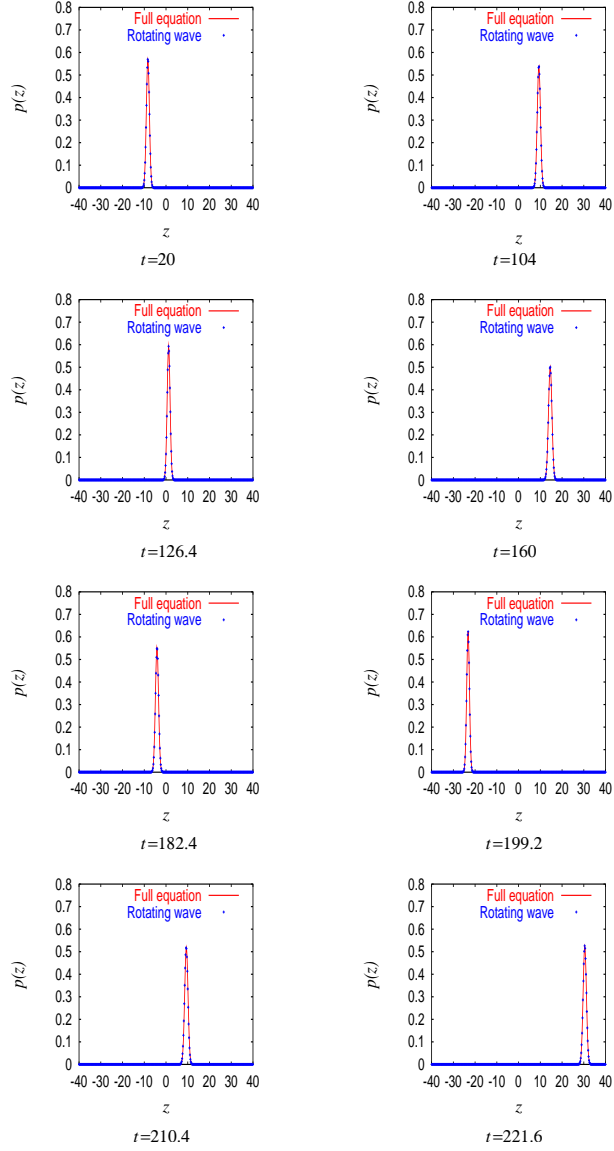,width=0.6\linewidth,angle=0}}
\caption{The probability distribution, $p(z)$, of finding the cantilever at
  position $z$ at a range of times
for the complete and rotating wave Hamiltonians.}
\label{fig:p_vs_z_approx}
\end{figure}

In this rotating-wave approximation, if the spin begins in
an instantaneous eigenstate of $\H_S(t)$, it will remain in an
instantaneous eigenstate at all times.  If it begins in a superposition
of the two eigenstates, the spin and cantilever degrees of freedom will
become entangled, with the two components of the wavefunction
corresponding to the two spin directions remaining undisturbed for all
times.  Monitoring the position of the cantilever then serves as a
nondemolition measurement of the spin, as we would wish.

Note that the corrections to the adiabatic approximation include terms
which can flip the spin.  These terms must remain small for the system
to be a true nondemolition measurement.  The result of the spin
measurement manifests itself as a $\pi$ phase shift in the oscillation of
the cantilever.  We can see this in figure 4.

\begin{figure}[htbp]
\centerline{\psfig{file=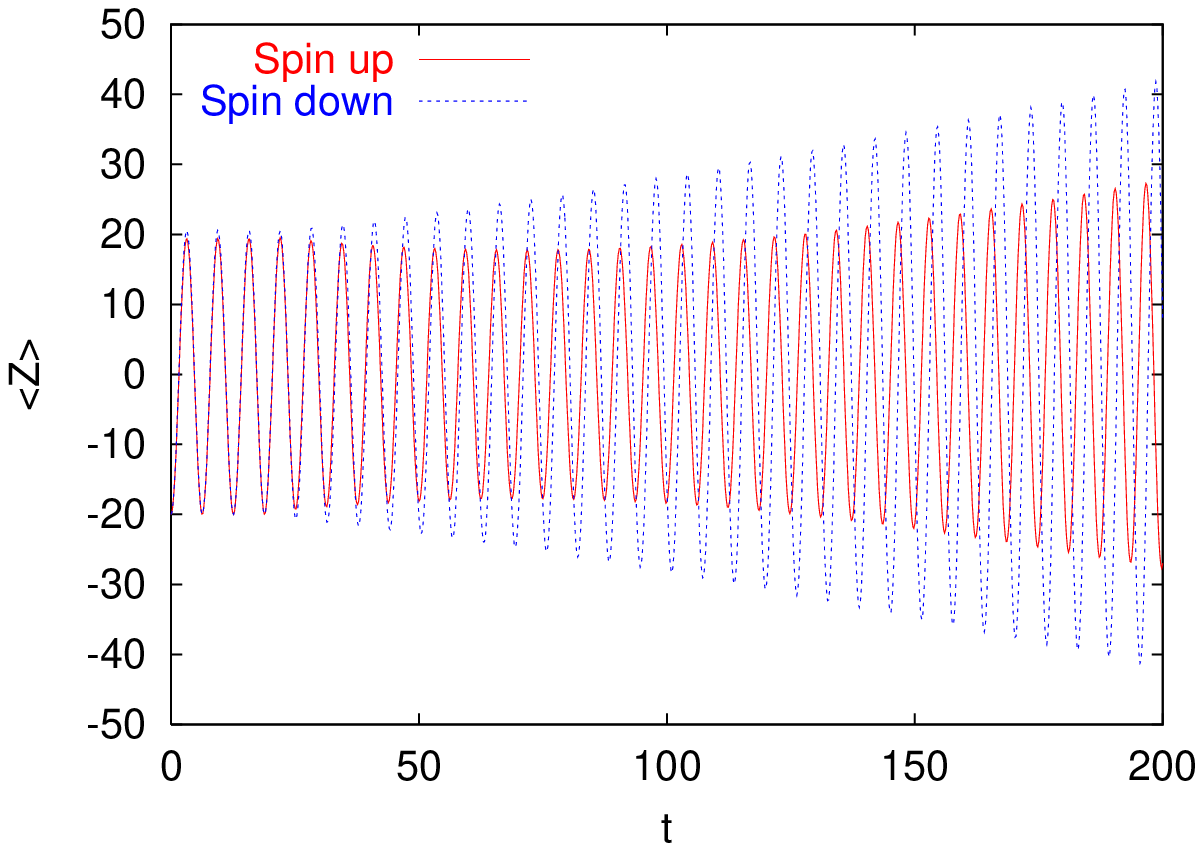,width=0.6\linewidth,angle=0}}
\centerline{\psfig{file=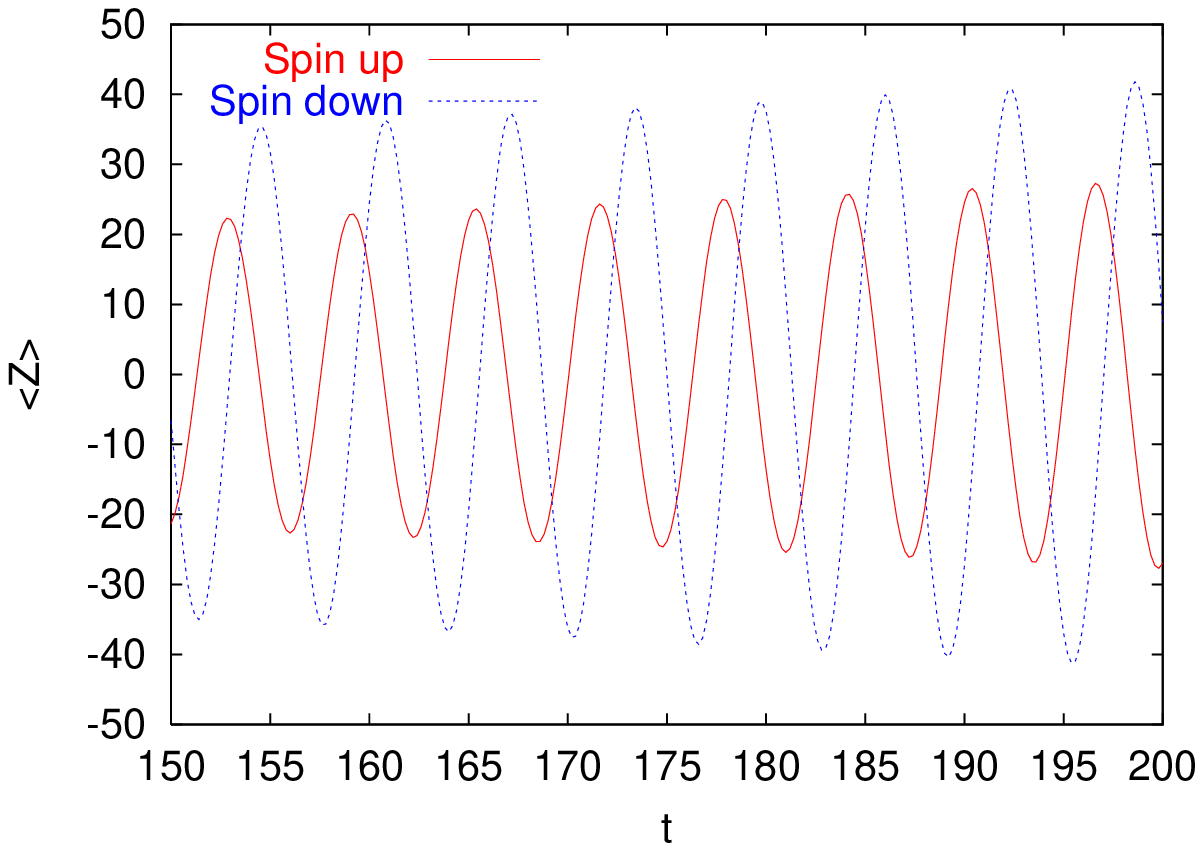,width=0.6\linewidth,angle=0}}
\caption{Mean cantilever position $\expect{\Z}$ vs. $t$ for initial
spin up and down in the $\Sz'$ direction.}
\label{fig:z_vs_t_phase}
\end{figure}

\section{The thermal environment}

Unfortunately, in practice we cannot treat the cantilever as an
isolated system.  It is coupled at least weakly to the vibrational modes
of the bulk, and is therefore subject to dissipation and thermal noise.
Since the cantilever can be treated as a single harmonic oscillator,
we can model the effects of this thermal bath by the well-known
Caldeira-Leggett \cite{Leggett83} master equation in the
high-temperature limit:
\begin{equation}
{\dot\rho} = - \frac{i}{\hbar} [\H_{SZ}(t),\rho]
  - \frac{i\gamma_m}{\hbar} [\Z,\{\p,\rho\}]
  - \frac{\gamma_m}{2\ell^2} [\Z,[\Z,\rho]] \;,
\label{CaldLeggEqn}
\end{equation}
where the parameters are
\begin{eqnarray}
\gamma_m &=& \frac{\Gamma}{2m} \;, \nonumber\\
\ell &=& \frac{\hbar}{2\sqrt{mkT}} \;,
\end{eqnarray}
$m$ is the cantilever mass, $T$ is the temperature, $k$ is Boltzmann's
constant (or the equivalent for our system of units), and $\Gamma$
is the strength of the coupling to the thermal bath.  We can interpret
$\gamma_m$ (with units of inverse time) as the dissipation rate and
$\ell$ (with units of length) as the thermal de~Broglie wavelength.

A feature of this equation is that it doesn't necessarily preserve the
positivity of $\rho$ on short time scales (though at long times it is
well-behaved) \cite{Hu92}.  
This arises because of the approximations which are
made in the derivation, which become invalid at very short times.  While
this may be physically unimportant, it can be inconvenient; in particular,
if we wish to {\it unravel} the evolution into a stochastic Schr\"odinger
equation \cite{Carmichael96} (as we will in section 6), it is
necessary to start with a 
master equation in {\it Lindblad form}, \cite{Lindblad76}
\begin{equation}
{\dot\rho} = - \frac{i}{\hbar}[\H,\rho]
  + \sum_j \left[ 2 \L_j \rho \Ldag_j - \{\Ldag_j\L_j,\rho\} \right] \;,
\label{LindbladEqn}
\end{equation}
for some Hermitian $\H$ and set of general {\it Lindblad operators} $\{\L_j\}$.
The Caldeira-Leggett equation (\ref{CaldLeggEqn}) is not of this form,
which is why it can violate positivity of $\rho$.

The exact quantum Brownian motion master equation 
was shown \cite{Hu92} not to have the Lindblad form, but rather requires
time-dependent coefficients to ensure the positivity of the density
matrix at short times.  However, by keeping more terms from the high- or
medium-temperature-limit expansion in a consistent way,
Di\'osi \cite{Diosi93} has shown that the Caldeira-Leggett
equation can be replaced 
by another master equation which {\it is} of Lindblad form, and which agrees
with it except at very short times when the equation's validity is
questionable in any case.  This is done by adding a term
to (\ref{CaldLeggEqn}) of the form
$-(\gamma_m\ell^2/2\hbar^2)[\p,[\p,\rho]]$.  The procedure is analogous
to completing the square.  If we choose the ansatz
\begin{equation}
\L = A \Z + i B \p
\end{equation}
with real $A,B$,
plug it into the equation (\ref{LindbladEqn}), and equate it to the
Caldeira-Leggett equation (\ref{CaldLeggEqn}) plus the additional term,
we get
\begin{eqnarray}
{\dot\rho} &=& - (i/\hbar)[\H,\rho]
  - A^2 [\Z,[\Z,\rho]] - B^2 [\p,[\p,\rho]] \nonumber\\
&& + i AB (- 2\Z\rho\p + \Z\p\rho + \rho\Z\p
  + 2\p\rho\Z - \p\Z\rho - \rho\p\Z) \nonumber\\
&=& -(i/\hbar)[\H_{SZ}(t),\rho]
  - \frac{\gamma_m}{2\ell^2} [\Z,[\Z,\rho]]
  - \frac{\gamma_m\ell^2}{2\hbar^2} [\p,[\p,\rho]] \nonumber\\
&& + \frac{i\gamma_m}{\hbar} (\p\rho\Z - \Z\p\rho -\Z\rho\p + \rho\p\Z) \;,
\end{eqnarray}
which implies that
\begin{eqnarray}
A &=& \sqrt{\gamma_m/2\ell^2} \;, \nonumber\\
B &=& \sqrt{\gamma_m\ell^2/2\hbar^2} \;, \nonumber\\
\H &=& \H_{SZ}(t) + (\gamma_m/2)(\Z\p + \p\Z) \equiv \H'_{SZ}(t) \;.
\end{eqnarray}
So the Lindblad operator for this equation is
\begin{equation}
\L = \sqrt{\gamma_m/2}\left( (1/\ell)\Z + i(\ell/\hbar)\p \right) \;,
\label{LindbladOp}
\end{equation}
and the effective Hamiltonian, going to the rotating picture and
making use of the approximation derived in section 3, is
\begin{equation}
\H'_{SZ}(t) = \frac{1}{2m}\p^2 + \frac{m\omega_m^2}{2}\Z^2
  - 2 \eta (f(t)/\lambda(t)) \Z \Sz'
  + (\gamma_m/2)(\Z\p + \p\Z) \;.
\label{Heff}
\end{equation}
In order for the cantilever to be an effective measurement device,
the loss rate must be very low:  $\omega_m \gg \gamma_m$.

\section{The effects of monitoring}

In order to serve as a measurement scheme, we must have some way of
{\it monitoring} the motion of the cantilever.  Because of the microscopic
scale of the motion, this is not so easily done.  One approach is to
use optical interferometry to measure the cantilever position.

As shown in figure 1, the cantilever forms one side of an optical
microcavity and the cleaved end of the fiber forms the other side.
As the cantilever moves, the resonant frequency of the
cavity changes.  Because the timescale of the cantilever's motion is
very long compared to the optical timescale, we can treat the effects
of this in the adiabatic limit.  The cavity mode is also subject to
driving by an external laser, and has a very high loss rate.
The full master equation \cite{Milburn94}
for the cantilever-spin-cavity system in
the interaction picture is
\begin{eqnarray}
{\dot\rho} &=& - \frac{i}{\hbar}[\H'_{SZ}(t),\rho]
  + 2\L\rho\Ldag - \Ldag\L\rho - \rho\Ldag\L \nonumber\\
&& - i [ E(\adag+\a) + \adag\a(\Delta + \kappa\Z),\rho] \nonumber\\
&& + (\gamma_c/2)(2\a\rho\adag - \adag\a\rho - \rho\adag\a) \;,
\label{full_master}
\end{eqnarray}
where $\H'_{SZ}(t)$ and $\L$ are the Hamiltonian and Lindblad
operator for the cantilever and spin given by eqns.~(\ref{LindbladOp})
and (\ref{Heff}), $E$ is the strength of the laser driving,
$\Delta$ is the detuning from the ``neutral'' cavity frequency,
$\kappa$ is the coupling strength of the cantilever to the cavity mode,
and $\gamma_c$ is the loss rate of the cavity.

Suppose now that we perform homodyne measurement \cite{Carmichael93,Wiseman93} 
on the light which
escapes from the cavity.  We would like to replace the equation
(\ref{full_master}) above with an equation for the {\it conditional
evolution} of $\rho$, conditioned on the output photocurrent $I_c(t)$.
The conditional evolution equation for our system then becomes
\cite{Wiseman93,Doherty99} 
(in It\^o calculus form)
\begin{eqnarray}
d\rho &=& - \frac{i}{\hbar}[\H'_{SZ}(t),\rho] dt
  + \left( 2\L\rho\Ldag - \Ldag\L\rho - \rho\Ldag\L \right) dt \nonumber\\
&& - i [ E(\adag+\a) + \adag\a(\Delta + \kappa\Z),\rho] dt \nonumber\\
&& + (\gamma_c/2)\left(2\a\rho\adag - \adag\a\rho - \rho\adag\a\right) dt
\nonumber\\
&& + \sqrt{\gamma_c e_d} \left( \a\rho + \rho\adag
  - \expect{\a+\adag}\rho \right) dW_t \;,
\label{partial_unravelling}
\end{eqnarray}
where $0 \le e_d \le 1$ is the detector efficiency and $dW_t$ is
a real stochastic differential variable which obeys the statistics
\begin{equation}
M[dW_t] = 0 \;, \ \ M[dW_t dW_s] = \delta(t-s) dt \;.
\end{equation}
This noise is related to the output photocurrent
\cite{Carmichael93,Wiseman93,Doherty99} 
\begin{equation}
I_c(t) = \beta \left[ \gamma_c e_d \expect{\a+\adag}_t
  + \sqrt{\gamma_c e_d} \frac{dW_t}{dt} \right] \;,
\label{Ic}
\end{equation}
where $\beta$ is a constant giving the device's range of response.

We want to operate in the ``bad cavity'' limit where $\gamma_c \gg \omega_m$.
This means that the
cavity mode will approach equilibrium on a timescale very short compared
to that of the cantilever's motion, so that the cavity mode can be
adiabatically eliminated 
\cite{Milburn94,Wiseman93,Doherty99}
from this equation, leaving an equation in
terms of the spin and cantilever position alone.

Let the detuning vanish $\Delta\rightarrow0$ and the coupling $\kappa$
to the cantilever be very small.  If we initially neglect this coupling
altogether, we can solve for the steady-state of the cavity mode
in isolation from the cantilever:
\begin{eqnarray}
- i [ E(\adag+\a),\rho]
  + (\gamma_c/2)\left(2\a\rho\adag - \adag\a\rho - \rho\adag\a\right) &=& 0 \;,
  \nonumber\\
\left( \a\rho + \rho\adag
  - \expect{\a+\adag}\rho \right) &=& 0 \;,
\end{eqnarray}
which implies that $\rho = \ket{\alpha_0}\bra{\alpha_0}$, where
$\a\ket{\alpha_0} = \alpha_0\ket{\alpha_0}$ is a coherent state with
\begin{equation}
\alpha_0 = - \frac{2iE}{\gamma_c} \;.
\end{equation}

Now let us restore the coupling $\kappa$ between the cantilever and the
cavity mode.  If this coupling is very small, then the state of the
cavity mode will remain very close to the state $\ket{\alpha_0}$.
In this case, it is very useful to switch to a {\it displaced basis}
\cite{Milburn94,Wiseman93,Doherty99}
for the cavity mode.  We switch from the operators $\a,\adag$ to
{\it displaced operators}
\begin{eqnarray}
\b \equiv \a - \alpha_0  \;, \nonumber\\
\bdag \equiv \adag - \alpha^*_0  \;,
\label{displaced_op}
\end{eqnarray}
and {\it displaced number states}
\begin{equation}
\bdag\b \ket{n} = n \ket{n} \;.
\end{equation}
Obviously $\ket{0} = \ket{\alpha_0}$, and
$\ket{1} = \adag\ket{\alpha_0}-\alpha_0^*\ket{\alpha_0}$.

We now make the ansatz of keeping the two lowest displaced number
states $\ket{0,1}$ of the cavity mode and neglecting the rest.
\cite{Wiseman93,Milburn94,Doherty99}
We then write the full density matrix for the spin-cantilever-cavity
system as
\begin{equation}
\rho(t) = \rho_0(t) \otimes \ket0\bra0 + \rho_1(t) \otimes \ket1\bra0
  + \rho_1^\dagger(t) \otimes \ket0\bra1 + \rho_2(t) \otimes \ket1\bra1 \;,
\label{cavity_ansatz}
\end{equation}
where $\rho_{0,1,2}$ are operators which act on the Hilbert space of
the cantilever and spin, and $\rho_{0,2}$ are self-adjoint.  The
reduced density matrix of the spin-cantilever alone is obtained by
tracing out the cavity mode, yielding
\begin{equation}
\rho_{SZ}(t) = \rho_0(t) + \rho_2(t) \;.
\end{equation}
If we substitute the definitions (\ref{displaced_op}) and
(\ref{cavity_ansatz}) into the stochastic master equation
(\ref{partial_unravelling}) and collect terms, we get a set of
coupled equations in the operators $\rho_{0,1,2}$:
\begin{eqnarray}
d\rho_0 &=& \left( - \frac{i}{\hbar}[\H'_{SZ}(t),\rho_0]
  + 2\L\rho_0\Ldag - \Ldag\L\rho_0 - \rho_0\Ldag\L \right) dt \nonumber\\
&& - \frac{4i\kappa E^2}{\gamma_c^2} [\Z,\rho_0] dt
  + \frac{2\kappa E}{\gamma_c} (\Z\rho_1 + \rho_1^\dagger\Z) dt
  + \gamma_c \rho_2 dt \nonumber\\
&& + \sqrt{\gamma_c e_d} \left( \rho_1 + \rho_1^\dagger
  - \rho_0 \tr\{\rho_1 + \rho_1^\dagger\} \right) dW_t \;,
\end{eqnarray}
\begin{eqnarray}
d\rho_1 &=& \left( - \frac{i}{\hbar}[\H'_{SZ}(t),\rho_1]
  + 2\L\rho_1\Ldag - \Ldag\L\rho_1 - \rho_1\Ldag\L \right) dt \nonumber\\
&& - i\kappa\Z\rho_1 dt - \frac{4i\kappa E^2}{\gamma_c^2} [\Z,\rho_1] dt
  - \frac{2\kappa E}{\gamma_c} (\Z\rho_0 - \rho_2\Z) dt
  - (\gamma_c/2) \rho_1 dt \nonumber\\
&& + \sqrt{\gamma_c e_d} \left( \rho_2
  - \rho_1 \tr\{\rho_1 + \rho_1^\dagger\} \right) dW_t \;,
\end{eqnarray}
\begin{eqnarray}
d\rho_2 &=& \left( - \frac{i}{\hbar}[\H'_{SZ}(t),\rho_2]
  + 2\L\rho_2\Ldag - \Ldag\L\rho_2 - \rho_2\Ldag\L \right) dt \nonumber\\
&& - \left( i\kappa + \frac{4i\kappa E^2}{\gamma_c^2}\right) [\Z,\rho_2] dt
  - \frac{2\kappa E}{\gamma_c} (\Z\rho_1^\dagger + \rho_1\Z) dt
  - \gamma_c \rho_2 dt \nonumber\\
&& - \sqrt{\gamma_c e_d} \rho_2 \tr\{\rho_1 + \rho_1^\dagger\} dW_t \;.
\end{eqnarray}

Both $\rho_1$ and $\rho_2$ contain damping terms, which imply that
they will remain small at all times provided $\kappa\Z$ is sufficiently
small compared to $\gamma_c$.  (This also implies that our ansatz is
reasonable for sufficiently small $\kappa$.)

By making use of the above equations, we can find the evolution
equation for the reduced density matrix $\rho_{SZ}$:
\begin{eqnarray}
d\rho_{SZ}(t) &=& d\rho_0(t) + d\rho_2(t) \nonumber\\
&=& \left( - \frac{i}{\hbar}[\H'_{SZ}(t),\rho_{SZ}]
  + 2\L\rho_{SZ}\Ldag - \Ldag\L\rho_{SZ}
  - \rho_{SZ}\Ldag\L \right) dt \nonumber\\
&& - \frac{4i\kappa E^2}{\gamma_c^2} [\Z,\rho_{SZ}] dt
  + \frac{2\kappa E}{\gamma_c} [\Z,\rho_1 - \rho_1^\dagger] dt
  - i\kappa [\Z,\rho_2] dt \nonumber\\
&& + \sqrt{\gamma_c e_d} \left( \rho_1 + \rho_1^\dagger
  - \rho_{SZ} \tr\{\rho_1 + \rho_1^\dagger\} \right) dW_t \;.
\label{effective_master1}
\end{eqnarray}
If we keep only terms to second order in $\kappa\Z$ we can neglect
the $\rho_2$ term.  This leaves only the terms proportional to
$\rho_1 \pm \rho_1^\dagger$, which we need only know to leading order
in $\kappa\Z$.  Provided (as we have already assumed)
that the cantilever moves slowly compared to the timescale set by
$\gamma_c$ and that $\kappa\Z$ can be treated as small, then to leading
order $d\rho_1$ vanishes; $\rho_1$ remains in an approximate equilibrium
state.  If we make use of this assumption we can (again to leading
order) solve for $\rho_1 \pm \rho_1^\dagger$:
\begin{eqnarray}
\rho_1 + \rho_1^\dagger &\approx&
  - \frac{4\kappa E}{\gamma_c^2} \{\Z,\rho_{SZ}\} \;, \nonumber\\
\rho_1 - \rho_1^\dagger &\approx&
  - \frac{4\kappa E}{\gamma_c^2} [\Z,\rho_{SZ}] \;,
\end{eqnarray}
which when inserted into (\ref{effective_master1}) gives us a closed
evolution equation for $\rho_{SZ}$:
\begin{eqnarray}
d\rho_{SZ}(t) &=&
  \left( - \frac{i}{\hbar}[\H'_{SZ}(t),\rho_{SZ}]
  + 2\L\rho_{SZ}\Ldag - \Ldag\L\rho_{SZ}
  - \rho_{SZ}\Ldag\L \right) dt \nonumber\\
&& - \frac{4i\kappa E^2}{\gamma_c^2} [\Z,\rho_{SZ}] dt
  - \frac{8\kappa^2 E^2}{\gamma_c^3} [\Z,[\Z,\rho_{SZ}]] dt \nonumber\\
&& + \sqrt{\gamma_c e_d} \frac{4\kappa E}{\gamma_c^2}
  \left( \Z\rho_{SZ} + \rho_{SZ}\Z
  - 2 \rho_{SZ} \tr\{\rho_{SZ}\} \right) dW_t \;.
\label{effective_master2}
\end{eqnarray}
(Note that we have absorbed a factor of $-1$ into $dW_t$.)

Examining the terms in (\ref{effective_master2}), we see that by
eliminating the cavity mode we get another effective term in the
Hamiltonian, and another Lindblad operator.  We can therefore write
this stochastic master equation in the form
\begin{eqnarray}
d\rho_{SZ}(t) &=&
  - \frac{i}{\hbar}[\H_{\rm eff}(t),\rho_{SZ}] dt
  + \sum_{j=1}^2 \left( 2\L_j\rho_{SZ}\Ldag_j - \Ldag_j\L_j\rho_{SZ}
  - \rho_{SZ}\Ldag_j\L_j \right) dt \nonumber\\
&& + \sqrt{e_d/2} \left( (\L_2 - \expect{\L_2})\rho_{SZ}
  + \rho_{SZ} (\L_2 - \expect{\L_2}) \right) dW_t \;,
\label{effective_master3}
\end{eqnarray}
where we now make the definitions
\begin{eqnarray}
\L_1 &=&
  \sqrt{\gamma_m/2}\left( (1/\ell)\Z + i(\ell/\hbar)\p \right) \nonumber\\
\L_2 &=& \sqrt{8\kappa^2 E^2/\gamma_c^3}\Z \;, \nonumber\\
\H_{\rm eff}(t) &=&
  \frac{1}{2m}\p^2 + \frac{m\omega_m^2}{2}\Z^2
  - 2 \eta (f(t)/\lambda(t)) \Z \Sz' \nonumber\\
&& + \frac{4\kappa E^2}{\gamma_c^2}\Z
  + (\gamma_m/2)(\Z\p + \p\Z) \;.
\label{Heff2}
\end{eqnarray}
Note that the term $4\kappa E^2\Z/\gamma_c^2$ is a constant force,
which just displaces the equilibrium position of the cantilever.  It
can be eliminated simply by changing the origin of $\Z$, and is in
any case small for reasonable values of the parameters.
The output from the homodyne measurement now corresponds to a
measurement of the cantilever position $\expect{\Z}$:
\begin{equation}
I_c(t) = \beta \left(- \frac{8e_d\kappa E}{\gamma_c} \expect{\Z}
  + \sqrt{\gamma_c e_d} \frac{dW_t}{dt} \right) \;.
\label{IcZ}
\end{equation}
As we shall see in the next section, we can further unravel this
stochastic master equation (\ref{effective_master3}) into a stochastic
{\it Schr\"odinger} equation for pure states.  This further unraveling
provides considerable improvement in numerical efficiency, though it
does not represent an actual measurement process.

\section{Pure state unraveling}

The stochastic master equation (\ref{effective_master3}) represents
the evolution of the cantilever-spin system, conditioned on the
photocurrent measurement record $I_c(t)$.  If we averaged over all
possible measurement records, the $dW_t$ terms would average to zero,
and we would be left with an ordinary deterministic master equation
for the cantilever and spin.  It is for this reason that the stochastic
master equation is therefore often referred to as an {\it unraveling} of
the average master equation.

For numerical purposes, it is often much easier to solve an equation
for a {\it pure state vector} rather than a density matrix
\cite{Schack95,Carmichael96}.  It is
therefore useful to unravel equation (\ref{effective_master3}) still
further to an equation which preserves pure states.  We do this
by introducing a second stochastic process.

First, let us idealize to perfect detector efficiency $e_d=1$.  We then
introduce the new master equation
\begin{eqnarray}
d\rho_{SZ}(t) &=&
  - \frac{i}{\hbar}[\H_{\rm eff}(t),\rho_{SZ}] dt
  + \sum_{j=1}^2 \left( 2\L_j\rho_{SZ}\Ldag_j - \Ldag_j\L_j\rho_{SZ}
  - \rho_{SZ}\Ldag_j\L_j \right) dt \nonumber\\
&& + \sum_{j=1}^2 \left( (\L_j - \expect{\L_j})\rho_{SZ}
  + \rho_{SZ} (\L_j - \expect{\L_j}) \right) dW_{jt} \;,
\label{effective_master4}
\end{eqnarray}
where the Hamiltonian and Lindblad operators are the same as in
(\ref{Heff2}) and we now have two independent noise processes represented
by stochastic differential variables $dW_{1t}$ and $dW_{2t}$
which satisfy
\begin{equation}
M[dW_{jt}] = 0 \;, \ \ dW_{it} dW_{js} = \delta(t-s)\delta_{ij} dt \;.
\end{equation}
If we take the mean of (\ref{effective_master4}) over $dW_{1t}$ we
recover equation (\ref{effective_master3}).  We can think of the
additional stochastic process as representing a fictitious additional
measurement, whose outcome we average over to recover the state which
is conditioned on the {\it actual} measurement.

However, equation (\ref{effective_master4}) has a great advantage over
(\ref{effective_master3}).  If $\rho_{SZ}$ is initially a pure state
$\rho_{SZ} = \ket{\psi_{SZ}}\bra{\psi_{SZ}}$, it will remain a pure
state at all times, the state of course depending on the stochastic
processes $W_1$ and $W_2$.  We can recover the solution of
(\ref{effective_master3}) by averaging
\begin{equation}
\rho_{SZ}(t) = M_{W_1}[ \ket{\psi_{SZ}(t)}\bra{\psi_{SZ}(t)} ] \;.
\end{equation}

It would be useful to replace equation (\ref{effective_master4}) with
an explicit evolution equation for $\ket{\psi_{SZ}}$ instead of
$\rho_{SZ}$.  This equation is the {\it quantum state diffusion equation
with real noise} \cite{Gisin84,QSD}:
\begin{eqnarray}
d\ket{\psi_{SZ}} &=& - \frac{i}{\hbar} \H_{\rm eff}(t) \ket{\psi_{SZ}} dt
  + \frac{1}{\sqrt2} \sum_j \left( 2 \expect{\Ldag_j}\L_j
  - \Ldag_j\L_j - |\expect{\L_j}|^2 \right) \ket{\psi_{SZ}} dt \nonumber\\
&& + \frac{1}{\sqrt2} \sum_j \left( \L_j - \expect{\L_j} \right)
  \ket{\psi_{SZ}} dW_{jt} \;.
\label{qsd}
\end{eqnarray}
The nonlinearity of this equation arises to preserve the norm.

\section{Numerical simulation}

We have simulated this system using the C++ quantum state diffusion
library \cite{Schack97} to numerically solve both the unitary
evolution with Hamiltonian (\ref{rotating_approx}) and the stochastic
equation (\ref{qsd}).  All of the figures in this paper were generated
using this software.

We chose our parameters based on those used by Berman et al.,
\cite{Berman01b}.  These values are (in arbitrary units):
\begin{eqnarray}
\hbar &=& \omega_m = m = 1 \;, \nonumber\\
\eta &=& 0.3 \;, \nonumber\\
\varepsilon &=& 400.0 \;, \nonumber\\
\gamma_m &=& \omega_m/Q = 0.00001 \;, \nonumber\\
k_B T &=& 10000.0 \;,
\end{eqnarray}
where $Q$ is the quality factor of the cantilever.  The driving force
$f(t)$ takes the form
\begin{equation}
f(t) = \cases{-6000 + 300t & if $0 \le t \le 20$;\cr
  1000\sin(t-20) & if $t > 20$.}
\end{equation}
If we make contact with physical values for actual cantilevers
used in experiments, we have $\omega_m \approx 10^5 {\rm s}^{-1}$
and $m \approx 10^{-12} {\rm kg}$.  The value of $k_B T$ above then
corresponds to a temperature of around $0.1 {\rm K}$, which is within
the bounds of experimental feasibility, though rather lower than the
temperatures used in current experiments (around 3K) \cite{Rugar01}.
Since $\eta = (g\mu/2)(\partial B_z/\partial Z)_0$, the value of $\eta$
corresponds to a field gradient of about $1.5\times10^7 {\rm T/m}$, which is
higher than current experiments by roughly two orders of magnitude
\cite{Rugar01}, but hopefully this too will improve with time.  The
cantilever would undergo displacements of about a nanometer.

Alternatively, rather than increasing the field gradient we could
achieve similar numbers by lowering the spring constant of the cantilever,
for instance by shrinking the mass of the cantilever.  Lowering the
mass by a factor of 100 has the same relative effect on $\eta$ as increasing
the field gradient by a factor of ten.

We then might ask about realistic parameters for the monitoring.
A typical cavity size $L$ is about a micrometer, with a laser frequency
of $\omega_c \approx 1.4\times10^{15} {\rm s}^{-1}$.  This cavity is generally
quite lossy; reasonable quality factors might be in the range
$Q_c\sim$10--100.  The parameter $E$ is a function of the laser power,
$E = \sqrt{P\gamma_c/\hbar\omega_c} = \sqrt{P/\hbar Q_c}$.  For
$P \sim 1 \mu{\rm W}$ and $Q_c \sim 100$ we have $E \sim 10^{13}{\rm s}^{-1}$.
The coupling between the cantilever and the cavity is given by a
geometric factor $\kappa = \omega_c/L \sim 1.4\times10^{21}
({\rm m}\cdot{\rm s})^{-1}$.  In arbitrary units, this gives coefficients
\begin{eqnarray}
\frac{8\kappa E}{\gamma_c} &=& 1.9\times10^3 \;, \nonumber\\
\frac{4\kappa E^2}{\gamma_c^2} &=& 7\times10^2 \;, \nonumber\\
\sqrt{\frac{8\kappa^2 E^2}{\gamma_c^3}} &=& 0.07 \;.
\end{eqnarray}
The first value is the multiplier in (\ref{IcZ}); the second gives
the equilibrium displacement of the cantilever; the third is the
coefficient of the Lindblad operator $\L_2$.


One question we can now easily address is how quickly the state of
the spin collapses onto eigenstates of $\Sz'$.  In figure 5 we plot
$\expect{\Sz'}$ for ten different trajectories.  We see that in all
ten cases the spin converged to $\pm 1/2$ quite quickly, before $t=80$.
\begin{figure}[htbp]
\centerline{\psfig{file=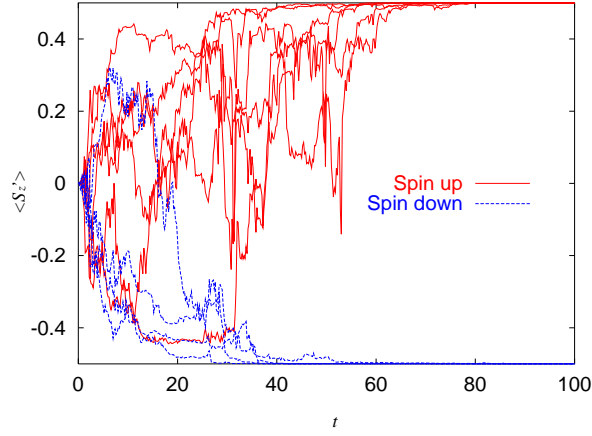,width=0.6\linewidth,angle=0}}
\caption{Expectation value $\expect{\Sz'}$ vs. t in arbitrary units
for ten different trajectories, showing the rapid localization of the
spin, for an initial superposition state
$\left(|v_+(0)\rangle+|v_-(0)\rangle\right)/\sqrt{2}$.}
\label{fig:sz_vs_t}
\end{figure}

If we compare this to the results of figure 4, we see that
the spin state collapses rather more quickly than the cantilever
oscillations can respond.  We only get a clear output signal when the
two phases are well separated, which does not occur until nearly
$t=150$.  Generically, the difficulty of collapsing the spin state
is much less than the difficulty of obtaining an unequivocal readout.

The curves depicted in figure 4 are idealized, without the measurement
noise which will always be present in the output current (\ref{Ic}) or
(\ref{IcZ}).  In figure 6 we show what actual output would look like for
the set of parameters we are discussing.  Note that even with the noise,
the two phases (representing spin up and spin down) are clearly
distinguishable.  In the next section, we derive an expression
for the signal-to-noise ratio in more general situations.

\begin{figure}[htbp]
\centerline{\psfig{file=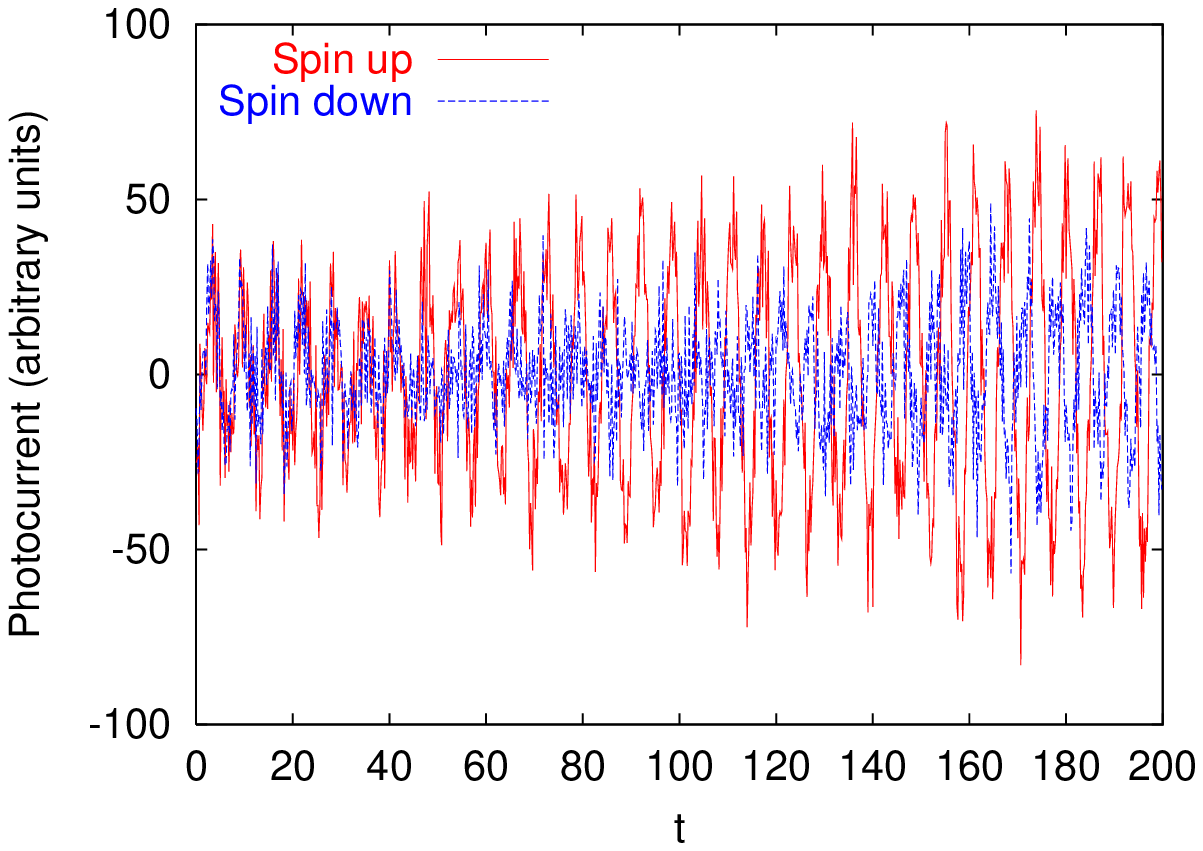,width=0.6\linewidth,angle=0}}
\centerline{\psfig{file=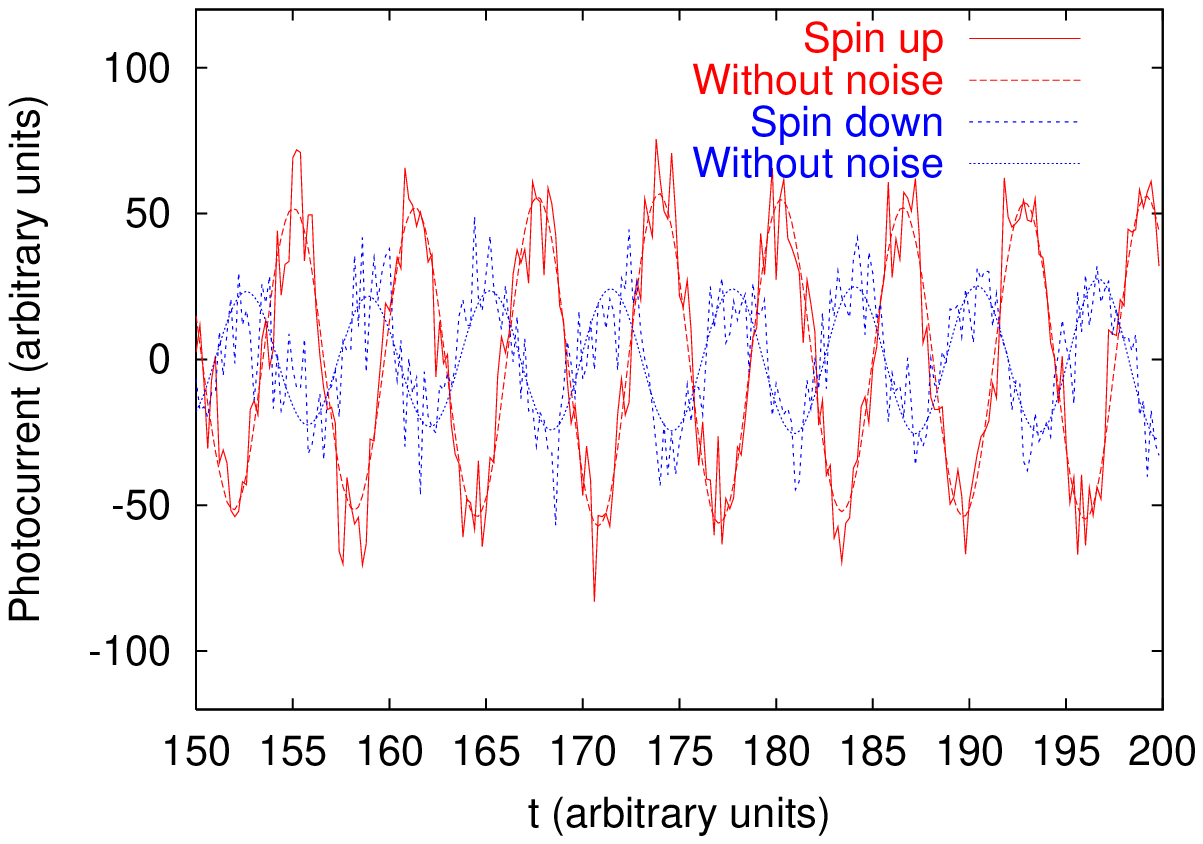,width=0.6\linewidth,angle=0}}
\caption{Simulation of photocurrent output in arbitrary units, including
measurement noise, using parameters of section 7.  We have chosen the
scale $\beta$ so that the vertical scale matches that of figure 4,
and also plotted the signals without the noisy $dW/dt$ components.}
\label{fig:simulated_output}
\end{figure}

\section{Signal-to-noise ratio}
\label{sec:SNR}

Since we have to detect the effect of very weak force on the
cantilever by the single spin, we need very high resolution for the
cantilever position measurements and a good control of
the various noise sources in the MRFM device. 
As described in section 2, the tiny 
displacement of the cantilever is measured by a  
fiber-optic interferometer as a phase shift of the
interference fringes.    
We shall analyze the quantum and thermal noise in this 
homodyne measurement scheme. 

The Hamiltonian for the combined system of the spin,
cantilever and cavity mode, excluding coupling to the environments,  
in the spin rotating frame is  
\begin{eqnarray}
  \label{eq:Htot}
  \hat{H}&=&\hat{H}_{Z}-2\eta\frac{f(t)}{\lambda (t)}\Z\Sz'
  +\hbar \omega_c \adag \a 
\nonumber \\
&& +\hbar E (\adag e^{-i\omega_0 t}+\a e^{i\omega_0 t})
   +\hbar \kappa \adag \a \hat{Z}.
\end{eqnarray}
Here $\omega_c$ is the optical frequency of the cavity mode, 
$\omega_0\sim \omega_c$ is the driving frequency of the external laser
and other terms and parameters have been described in section 5.
The master equation approach in section 4 is valid in high or medium 
temperature case. Here we analyze the noise in the Heisenberg picture,
using the quantum Langevin equation approach that is valid at any
temperature. \cite{Vitali01} 

Using standard techniques, \cite{Walls94,Gardiner00} 
the reservoir (environmental) variables may
be eliminated, in the interaction picture with respect
to $\hbar \omega_0 \adag \a $, to give 
the quantum Langevin equations describing the dynamics of the whole system: 
\begin{eqnarray}
  \label{eq:Langevin}
 \frac{d\Z(t)}{dt}&=&\frac{1}{m}\p(t),
\\
\frac{d\p(t)}{dt}&=&-m\omega_m^2\Z(t)-\frac{\Gamma}{m}\p(t)-\hbar\kappa
\adag(t) \a(t)+\hat{{\cal W}}(t)
+2\eta\frac{f(t)}{\lambda (t)}\Sz'(t),
\\
\frac{d\a(t)}{dt}&=&-(i\omega_c-i\omega_0+\frac{\gamma_c}{2})\a(t)
-i\kappa
\Z(t)\a(t)-iE+\sqrt{\gamma_c}\a_{\rm in}(t), 
\\
\frac{d\Sz'(t)}{dt}&=&0,
\\
\frac{d\Sx'(t)}{dt}&=&2\eta\frac{f(t)}{\lambda (t)}\Z(t)\Sy'(t),
\\
\frac{d\Sy'(t)}{dt}&=&-2\eta\frac{f(t)}{\lambda (t)}\Z(t)\Sx'(t).
\end{eqnarray}
In the equations, the usual optical input noise 
operator $\a_{\rm in}(t)$ is associated
with the vacuum fluctuations of the continuum of electromagnetic modes
outside the cavity and its correlation function is given by
\begin{equation}
  \label{eq:ain}
  \langle\a_{\rm in}(t)\adag_{\rm in}(t')\rangle=\delta(t-t').
\end{equation}
The random force $\hat{\cal W}(t)$ 
describes the thermal noise motion 
(quantum Brownian motion) of the cantilever at temperature $T$. 
For the case of an Ohmic environment, 
the thermal random force correlation is given by  \cite{Vitali01}  
\begin{equation}
  \label{eq:Wcf}
\langle\hat{\cal W}(t)\hat{\cal W}(t')\rangle=\frac{\hbar\Gamma}{\pi}
[{\cal F}_r(t-t')+i{\cal F}_i(t-t')],  
\end{equation}
where 
\begin{eqnarray}
  \label{eq:Fr}
  {\cal F}_r(t)&=&\int_0^\Omega d\omega\, \omega \cos(\omega t)
\coth\left( \frac{\hbar \omega}{2 k_B T}\right),
\\
{\cal F}_i(t)&=&\int_0^\Omega d\omega\, \omega \sin(\omega t),
  \label{eq:Fi}
\end{eqnarray}
with $\Omega$ the frequency cutoff of the reservoir spectrum. 
Without the presence of the external driving force from the spin, the
cantilever-cavity system can be characterized by a semi-classical
steady state with a new equilibrium position for the cantilever,
displaced by $Z_{st}=-\kappa|\alpha_{st}|^2/(m\omega_m^2)$ with respect
to that with no external driving laser field, and 
the cavity mode in a coherent state $|\alpha_{st}\rangle$ with
the amplitude given by 
\begin{equation}
  \label{eq:a_st}
  \alpha_{st}=\frac{-iE}{\gamma_c/2+i\Delta},
\end{equation}
where $\Delta=\omega_c-\omega_0-\kappa^2|\alpha_{st}|^2/(m\omega_m^2)$
is the cavity mode detuning.
By adjusting either $\omega_0$ or $\omega_c$, the detuning can be set
to zero $\Delta=0$. As a result, $\alpha_{st}=\alpha_0=-2iE/\gamma_c$.
Linearizing the quantum Langevin equations about the steady-state
values and renaming with $\Z(t)$, $\a(t)$ the operators describing the
quantum fluctuations around the classical steady state,  
we obtain 
\begin{eqnarray}
  \label{eq:linearizedZ}
 \frac{d\Z(t)}{dt}&=&\frac{1}{m}\p(t),
\\
\frac{d\p(t)}{dt}&=&-m\omega_m^2\Z(t)-\frac{\Gamma}{m}\p(t)-\hbar\kappa
[\alpha_0\adag(t) +\alpha_0^*\a(t)]
\nonumber \\
&&+\hat{{\cal W}}(t) + 2\eta\frac{f(t)}{\lambda (t)}\Sz'(t),
\\
\frac{d\a(t)}{dt}&=&-\frac{\gamma_c}{2}\a(t)
-i\kappa \alpha_0 
\Z(t)+\sqrt{\gamma_c}\a_{\rm in}(t), 
  \label{eq:linearizeda}
\\
\frac{d\Sz'(t)}{dt}&=&0,
\\
\frac{d\Sx'(t)}{dt}&=&2\eta\frac{f(t)}{\lambda (t)} [Z_{st}+\Z(t)]\Sy'(t),
\\
\frac{d\Sy'(t)}{dt}&=&-2\eta\frac{f(t)}{\lambda (t)} [Z_{st}+\Z(t)]\Sx'(t).
\end{eqnarray}
In the bad cavity limit where $\gamma_c\gg \omega_m, (\Gamma/m), \kappa\Z$
[i.e., set $(d\a(t)/dt)=0$ in (\ref{eq:linearizeda})], the
dynamics of the field quadrature, $\adag(t)+\a(t)$,  
adiabatically follows
that of the cantilever position:
\begin{equation}
  \label{eq:X}
  \adag(t)+\a(t)=-i\frac{4\kappa\alpha_0}{\gamma_c}\Z(t)
+\frac{2}{\sqrt{\gamma_c}} [ \a_{\rm in}(t)+\adag_{\rm in}(t)].
\end{equation}
Thus monitoring this field quadrature
of the cavity mode via homodyne measurement corresponds to a
measurement of the cantilever position and hence the state of the spin. 
Using the usual input-output relation, \cite{Walls94,Gardiner00}
\begin{equation}
  \label{eq:in-out}
  \a_{\rm out}(t)=\sqrt{\gamma_c} \a(t)-\a_{\rm in},
\end{equation}
we may define an operator corresponding to the output current 
\begin{eqnarray}
\hat{I}_{\rm out}(t)& =& \sqrt{\gamma_c}\beta
[\a_{\rm out}(t)+\adag_{\rm out}(t)]
\nonumber \\
&=&\beta\{\gamma_c [\a(t)+\adag(t)]
- \sqrt{\gamma_c} [ \a_{\rm in}(t)+\adag_{\rm in}(t)]\}.       
  \label{eq:Iout}
\end{eqnarray}
Equation (\ref{eq:Iout}) is similar to (\ref{Ic}) with $e_d=1$.
By substituting  (\ref{eq:X}) into (\ref{eq:Iout}),
the resultant output current in the bad cavity limit is given by 
\begin{equation}
  \label{eq:Iz}
\hat{I}_{\rm out}(t) = \beta \left( \frac{-8\kappa E}{\gamma_c} \Z(t)
+ \sqrt{\gamma_c}\, [ \a_{\rm in}(t)+\adag_{\rm in}(t)]       
    \right) \;.  
\end{equation}
This equation is also similar to 
(\ref{IcZ}) with $e_d=1$, obtained from master equation approach.

The Langevin equations for $\Sx'$ and $\Sy'$ effectively decouple
from the other equations, since they do not appear on the
right-hand-side of the equations for the other variables.
Because of this, they have no effect in our estimate of
the signal-to-noise ratio, and we shall drop them henceforth.
Taking a Fourier transform of the linearized Langevin equations, 
we find, from (\ref{eq:Iout}), the Fourier component
of the output current as  
\begin{eqnarray}
  \label{eq:Iw}
\hat{I}_{\rm out}(\omega)&=&-\frac{\beta \gamma_c}{(i\omega-\gamma_c/2)}
\Biggl\{ \left( i\omega+\frac{\gamma_c}{2} \right) 
[ \a_{\rm in}(\omega)+\adag_{\rm in}(\omega)] \nonumber\\
&& -\frac{2i\kappa \alpha_0\sqrt{\gamma_c}}{m
  (\omega_m^2-\omega^2-i\Gamma\omega/m)}
\Biggl[ \frac{\hbar\kappa \alpha_0\sqrt{\gamma_c}}{(i\omega-\gamma_c/2)}
[ \adag_{\rm in}(\omega)-\a_{\rm in}(\omega)] \nonumber\\
&& \qquad\qquad\qquad +\hat{\cal W}(\omega)+G(\omega)\Sz']
\Biggr] \Biggr\}, 
\end{eqnarray}
where $G(\omega)$ is the Fourier transform 
of $G(t)=2\eta{f(t)}/{\lambda (t)}$.
The Fourier component of the mean output current signal is then given by
\begin{equation}
  \label{eq:Imean}
  |\langle \hat{I}_{\rm out}(\omega)\rangle|
=\beta\gamma_c \left(\frac{2\kappa\sqrt{\gamma_c}|\alpha_0|}{m}
\right)
\frac{|G(\omega)|}{|D(\omega)|}\langle\Sz'\rangle,
\end{equation}
where 
\begin{equation}
  \label{eq:Dw}
  D(\omega)=\left(i\omega-\frac{\gamma_c}{2}\right)
\left(\omega_m^2-\omega^2-i\omega\frac{\Gamma}{m}\right).
\end{equation}

The output current noise power density spectrum is defined as 
\begin{eqnarray}
  \label{eq:Sout}
  S_{\rm out}(\omega)&=&\left\{ \int d\tau\, e^{i\omega \tau}
\langle \hat{I}_{\rm out}(t)
  \hat{I}_{\rm out}(t+\tau)\rangle_{G(t)=0}\right\}_t
\nonumber\\
&=&\frac{1}{2\pi}\left\{ \int d\omega' e^{-i(\omega+\omega')t} 
 \langle \hat{I}_{\rm out}(\omega')
  I_{\rm out}(\omega)\rangle_{G(\omega)=0}\right\}_t,
\end{eqnarray}
where the subscript $G(t)=0$ means evaluation in the absence of the
external driving force from the spin and 
$\{\cdots\}_t$ denotes the time average over $t$.
To calculate this noise spectrum,
the Fourier transform of the noise correlation functions 
(\ref{eq:ain})--(\ref{eq:Fi}) is needed and given by 
\begin{eqnarray}
  \label{eq:ainw}
\langle\a_{\rm in}(\omega)\adag_{\rm in}(\omega')\rangle
&=&2\pi\delta(\omega+\omega'),
\\  
\langle\hat{\cal W}(\omega)\hat{\cal W}(\omega')\rangle
&=&2\pi\hbar \Gamma\omega\left[ 
1+\coth\left(\frac{\hbar\omega}{2k_BT}\right) \right]
\delta(\omega+\omega'),
\label{eq:Wcfw}
\end{eqnarray}
where in obtaining (\ref{eq:Wcfw}) the infinite frequency cutoff limit
of the Ohmic thermal reservoir spectrum, $\Omega\to \infty$, has been
assumed. After some calculations, one can then obtain the output noise
spectrum as
\begin{eqnarray}
  S_{\rm out}(\omega)&=&\beta^2\gamma_c^2
\left\{1+4\left(\frac{\hbar\kappa^2\gamma_c|\alpha_0|^2}{m}\right)^2
\frac{1}{[(\gamma_c/2)^2+\omega^2]|D(\omega)|^2}\right.
\nonumber\\
&&\left. \quad +4\left( \frac{\kappa^2\gamma_c|\alpha_0|^2\Gamma}{m^2}\right)
\frac{\hbar \omega}{|D(\omega)|^2}
\coth\left(\frac{\hbar\omega}{2k_BT}\right)\right\}.
  \label{eq:Soutw}
\end{eqnarray}
The first term in (\ref{eq:Soutw}), independent of frequency, is the
contribution from the shot noise of the photons. The next term is
the ``back-action'' noise on the position of the cantilever by the
radiation (photons).
This back action is due to the random way in which photons bounce off
the cantilever. The final term is the thermal noise,
due to the thermal Brownian-motion fluctuation of
the cantilever.
Equation (\ref{eq:Soutw}) is valid at all
temperatures.
The assumptions made in its derivation are the linearization around the
semi-classical steady state and the infinite frequency cutoff
$\Omega\to\infty$.
The high (or medium) temperature limit $\hbar\omega_m\ll k_BT$ 
can be obtained by approximating
\begin{equation}
  \label{eq:approx}
  \hbar\omega \coth\left(\frac{\hbar\omega}{2k_BT}\right)
\approx 2k_B T+\frac{\hbar^2\omega^2}{6k_B T}.
\end{equation}
We plot these three contributions to the noise for the simulation
parameters given in section 7.  We see that at the oscillator resonance
$\omega_m = 1$, thermal noise dominates.

\begin{figure}[htbp]
\centerline{\psfig{file=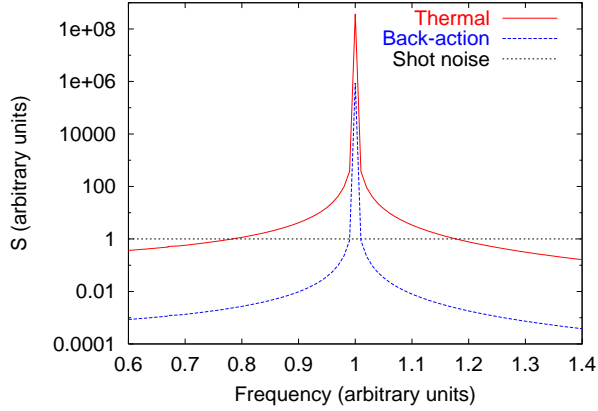,width=0.6\linewidth,angle=0}}
\caption{We plot the various terms of $S_{\rm out}(\omega)$ vs. $\omega$,
using the parameters of section 7.  We have scaled $\gamma_c\beta=1$.
Note that at $\omega=\omega_m=1$ the thermal noise dominates.}
\label{fig:noise}
\end{figure}

Let us define the signal-to-noise ratio per root Hertz as 
\begin{equation}
  \label{eq:SNR}
{\rm SNR}(\omega)=\frac{|\langle \hat{I}_{\rm out}(\omega)\rangle|}
{\sqrt{S_{\rm out}(\omega)}}.   
\end{equation}
We are interested in evaluating SNR$(\omega)$ at
the frequency equal to the cantilever vibration
frequency, $\omega=\omega_m$.
Note that 
\begin{equation}
  \label{eq:Dwm}
  \frac{1}{|D(\omega_m)|}=
\frac{1}{[(\gamma_c/2)^2+\omega_m^2]^{1/2}} \left(\frac{Q}{\omega_m^2}\right),
\end{equation}
where the quality factor $Q={m\omega_m}/{\Gamma}$.
As a result, the mean output current signal (\ref{eq:Imean})
at $\omega=\omega_m$
is enhanced by a factor of $Q\gamma_c/[(\gamma_c/2)^2+\omega_m^2]^{1/2}$
as compared with the $\omega=0$ case. However, a similar enhancement
occurs in the back-action noise
and the thermal noise terms. In other words, driving the cantilever at
$\omega=\omega_m$ amplifies not only its vibration
amplitudes due to the the driving force, but also the noise amplitude
due to the back-action radiation pressure and thermal Brownian motion
(see Fig.\ \ref{fig:noise}). 
We find ${\rm SNR}(\omega=\omega_m)$ can be written as
\begin{equation}
  \label{eq:SNRwm}
{\rm SNR}(\omega_m)
=\frac{|G(\omega_m)|\langle\Sz'\rangle}{\sqrt{N(\omega_m)}}
\end{equation}
where 
\begin{equation}
  \label{eq:Nwm}
N(\omega_m)= \frac{[(\gamma_c/2)^2+\omega_m^2]}{4\kappa^2\gamma_c|\alpha_0|^2}
\left(\frac{m\omega_m^2}{Q}\right)^2
  + \frac{\hbar^2\kappa^2\gamma_c|\alpha_0|^2}
  {[(\gamma_c/2)^2+\omega_m^2]}+\Gamma\hbar\omega_m 
\coth\left(\frac{\hbar\omega_m}{2k_BT}\right).
\end{equation}
We may set $\langle\Sz'\rangle=\pm(1/2)$ to estimate the
signal-to-noise ratio per root Hz, corresponding respectively to
the spin in the two different states in the rotating frame.

Because the driving force $f(t)$ is periodic, $G(\omega)$ is equal to
a sum of delta functions at $\omega=\omega_m, 3\omega_m, 5\omega_m, \ldots$.
Averaging over a small interval about $\omega_m$,
we can integrate over the delta function to get a value (for our
simulation parameters) of ${\rm SNR}(\omega_m) \approx 220 {\rm s}^{-1/2}$.
Thus, given a bandwidth of about 1Hz, this should be easily detectable
by our measurement scheme.  As mentioned in section 7, we have
assumed a magnetic field gradient roughly two orders of magnitude greater
than current experiments, and a much lower temperature.
A single spin, therefore, would be below the edge of detectability
by current experimental techniques.  Steady improvement in the field
strength, temperature and spring constant of these experiments,
however, should soon make single-spin measurement possible.

If the dominant noise source in MRFM comes from the thermal Brownian
motion of the cantilever, we can estimate the minimum detectable force
(when the signal-to-noise ratio is one) 
by keeping only the last term of (\ref{eq:Nwm}).  
In this case, with a measurement bandwith $\Delta \nu$, we obtain 
from (\ref{eq:SNRwm}), (\ref{eq:Nwm}) and (\ref{eq:approx}) 
the usual expression of the minimum detectable force at the high-temperature
limit ($\hbar\omega_m\ll k_BT$) as
\begin{equation}
\label{eq:Fmin}
F_{\min}=\sqrt{N(\omega_m)\Delta\nu}
=\sqrt{\frac{2kk_BT\Delta\nu}{Q\omega_m}}\; ,
\end{equation}
where $k=m\omega_m^2$ is the spring constant of the cantilever.  We
see, then, that improvement can come either from raising the force
(by increasing the field gradient), lowering the temperature, or lowering
the spring constant.

\section{Conclusion}

We have derived an approximate description of single-spin measurement by
magnetic resonance force microscopy, including both thermal noise and
measurement back-action, and used it to produce numerical simulations 
of a single-spin measurement.  These simulations use the quantum
trajectory method for open quantum systems.  The parameters we assumed for
this simulation were somewhat optimistic; but given the steady improvement in
experimental technique, we believe that measurements of this type will
be possible in the near future.

Single-spin measurements would be very useful in the construction of
solid-state quantum computers, in which the spin of an electron represents
a single qubit of information.  Given the great interest in solid-state
implementations as a possibly scalable realization of quantum computers,
finding practical ways to measure single spins would be very useful.
The results of our simulations suggest that magnetic resonance force
microscopy is a very promising approach to this difficult problem.

\section*{Acknowledgment}
\label{sec:Acknow}

HSG would like to thank G.~P.~Berman, G.~J.~Milburn, D.~V.~Pelekhov 
and P.~C.~Hammel 
for useful discussions. HSG would also like to acknowledge
support from the Hewlett-Packard Fellowship.  TAB was supported
by the Martin A.~and Helen Chooljian Membership in Natural Sciences,
and DOE Grant No.~DE-FG02-90ER40542.


\begin{thebibliography}{99}
%
\bibitem{Loss98}
D. Loss and D.~P. DiVincenzo,
Phys. Rev. A {\bf 57}, 120 (1998).

\bibitem{Kane98} 
B.E. Kane, Nature 
{\bf 393},  133-137  (1998). 

\bibitem{Yablonovitch99}
R. Vrijen, E. Yablonovitch, K. Wang, H.~W. Jiang, A. Balandin, 
V. Roychowdhury, T. Mor, and D. DiVincenzo 
Phys. Rev. A {\bf 62}, 012306 (2000).

\bibitem{Berman00a}
G.P. Berman, G.D. Doolen, P.C. Hammel, and V.I. Tsifrinovich, 
Phys. Rev. B {\bf 61}, 14694 (2000).

\bibitem{Twamley02}
J.~Twamley, quant-ph/0210202.

\bibitem{Loss02}
H.-A. Engel and D. Loss
Phys. Rev. B {\bf 65}, 195321 (2002).

\bibitem{Sidles91}
J.A. Sidles, 
 Appl. Phys. Lett. {\bf 58}, 2854 (1991).

\bibitem{Sidles92}
J.A. Sidles, 
Phys. Rev. Lett. {\bf 68}, 1124 (1992).

\bibitem{Berman01b}
G.P Berman, F. Borgonovi, G. Chapline, S.A. Gurvitz, P.C. Hammel,
D.V. Pelekhov, A. Suter and V.I. Tsifrinovich, quant-ph/0108025.

\bibitem{Rugar98}
K. Wago, D. Botkin, C.S. Yannoni, and D. Rugar, 
Phys. Rev. B {\bf 57}, 1108 (1998). 

\bibitem{Rugar01}
B.~C. Stipe, H.~J. Mamin, C.~S. Yannoni,  
T.~D. Stowe, T.~W. Kenny, and D. Rugar, 
Phys. Rev. Lett. {\bf 87}, 277602 (2001).

\bibitem{Rugar94}
D. Rugar, O. Z\"uger, S. Hoen, C.S. Yannoni, H.M. Vieth, and R.D. Kendrick, 
Science {\bf 264}, 1560  (1994).

\bibitem{Berman02}
G.P. Berman, F. Borgonovi, H.-S. Goan, S.A. Gurvitz and
V.I. Tsifrinovich, quant-ph/0210043, to appear in Phys. Rev. B.

\bibitem{Leggett83}
A.O. Caldeira and A.J. Leggett, Physica A {\bf 121}, 587 (1983).

\bibitem{Carmichael93} H.~J.~Carmichael, {\it An Open System Approach
        to Quantum Optics}, Lecture notes in physics (Springer-Verlag,
        Berlin, 1993).

\bibitem{Lindblad76} G.~Lindblad, Commun. Math. Phys.
         {\bf 48}, 199 (1976).

\bibitem{Hu92} B.-L. Hu, J.P. Paz and Y. Zhang,
Phys. Rev. D {\bf 45}, 2843 (1992).

\bibitem{Diosi93}L. Di\'osi, Europhys. Lett. {\bf 22}, 1 (1993);
Physica A {\bf 199}, 517 (1993).

\bibitem{Milburn94} G.~J.~Milburn, K.~Jacobs and D.~F.~Walls
        Phys. Rev. A {\bf 50}, 5256 (1994).

\bibitem{Wiseman93}  H.~M.~Wiseman and G.J.~Milburn,
        Phys. Rev. A {\bf 47}, 642 (1993); {\bf 47}, 1652 (1993).

\bibitem{Doherty99}  A.~C.~Doherty and K.~Jacobs,
        Phys. Rev. A {\bf 60}, 2700 (1999).

\bibitem{Schack95}  R.~Schack, T.~A.~Brun, I.~C.~Percival,
        J. Phys. A. {\bf 28}, 5401 (1995).

\bibitem{Carmichael96} See, for example, H.~J.~Carmichael (guest Ed), Quantum
        Semiclass. Opt. {\bf 8}, 47-314 (1996) special issue and
        references therein.

\bibitem{Gisin84}  N.~Gisin, Phys. Rev. Lett. {\bf 52}, 1657 (1984).

\bibitem{QSD}  N.~Gisin and I.~C.~Percival, J. Phys. A {\bf 25},
  5677 (1992); {\it ibid.} {\bf 26}, 2245 (1993).

\bibitem{Schack97}
R.~Schack and T.~A.~Brun, Comp. Phys. Comm. {\bf 102}, 210 (1997).

\bibitem{Vitali01}
V. Giovannetti and D. Vitali, Phys. Rev. A {\bf 63}, 023812 (2001). 

\bibitem{Walls94}  D.~F.~Walls and G.~J.~Milburn, {\em Quantum Optics},
        (Springer-Verlag, Berlin, 1994).

\bibitem{Gardiner00}  C.~W.~Gardiner and P.~Zoller, {\em Quantum Noise},
        2nd ed.\ (Springer-Verlag, Berlin, 2000).

\end{thebibliography}
\end{document}